\documentclass{ifacconf}
\usepackage{graphicx}      
\usepackage{natbib}        
\usepackage{amsmath,amssymb,amsfonts}
\usepackage{algorithmic}
\usepackage{textcomp}
\usepackage{xcolor}
\usepackage{colortbl}
\usepackage{optidef}
\usepackage{bm}
\usepackage{verbatim}

\newcommand{\ba}{\begin{array}}
	\newcommand{\ea}{\end{array}}
\newcommand{\babc}{\begin{abc}}
	\newcommand{\eabc}{\end{abc}}
\newcommand{\bc}{\begin{center}}
	\newcommand{\ec}{\end{center}}
\newcommand{\be}{\begin{equation}}
\newcommand{\ee}{\end{equation}}
\newcommand{\bea}{\begin{eqnarray}}
\newcommand{\eea}{\end{eqnarray}}
\newcommand{\beas}{\begin{eqnarray*}}
	\newcommand{\eeas}{\end{eqnarray*}}
\newcommand{\bh}{\begin{hangitem}}
	\newcommand{\eh}{\end{hangitem}}
\newcommand{\bhi}{\begin{hangitem}}
	\newcommand{\ehi}{\end{hangitem}}
\newcommand{\bi}{\begin{itemize}}
	\newcommand{\ei}{\end{itemize}}
\newcommand{\bn}{\begin{enumerate}}
	\newcommand{\en}{\end{enumerate}}
\newcommand{\bq}{\begin{quote}}
	\newcommand{\eq}{\end{quote}}
\newcommand{\btb}{\begin{tabular}}
	\newcommand{\etb}{\end{tabular}}
\def\litem[#1]{\item[#1\hfill]}         
\newtheorem{remark}{Remark}

\def\bff{{\bf f}}
\def\bfg{{\bf g}}
\def\bfh{{\bf h}}

\def\bfl{{\bf l}}

\def\bfx{{\bf x}}

\def\bfz{{\bf z}}
\def\balpha{\mbox{\boldmath${\alpha}$}}

\def\btheta{\mbox{\boldmath${\theta}$}}

\def\bxi{\mbox{\boldmath${\xi}$}}

\def\CN{{\mathcal N}}

\def\CR{{\mathcal R}}

\usepackage{csquotes}
\usepackage{graphicx,pstool}
\usepackage{multirow}
\usepackage{float}
\usepackage{color}
\usepackage{array}

\begin{document}
\begin{frontmatter}

\title{Event-triggered Hybrid Energy-aware Scheduling in Manufacturing Systems}


\author{Zhean Shao}, 
\author{Wen Li}, and
\author{Ying Tan}

\address{University of Melbourne, 
   Melbourne, VIC, Australia \\
   (e-mail: zheans@student.unimelb.edu.au, \\
   \{wen.li3, yingt\}@unimelb.edu.au, 
   )}

\begin{abstract}                
Incorporating renewable energy sources (RESs) into manufacturing systems has been an active research area in order to address many challenges originating from the unpredictable nature of RESs such as photovoltaics.
In the energy-aware scheduling for manufacturing systems, the traditional off-line scheduling techniques cannot always work well due to their lack of robustness with respect to uncertainties coming from imprecise models or unexpected situations. On the other hand, on-line scheduling or rescheduling, which can improve the robustness by using the model and the latest measurements simultaneously, suffer from a high computational cost. 
This work proposes a hybrid scheduling framework, which combines the advantages of both off-line scheduling and on-line scheduling, to provide a balanced solution between robustness and computational cost. 
A novel concept of partially-dispatchable state is introduced. It can be treated as a constant in scheduling when the model works well. When the model does not work well, it is triggered as the variable to tune to improve the performance.
Such an event-triggered structure can reduce the number of rescheduling and computational costs while achieving a reasonable performance and enhancing system robustness.
Moreover, the choice of partially-dispatchable state also provides an extra design freedom in achieving green manufacturing.
Simulation examples on a manufacturing system, of which consists a 100-kW solar photovoltaic system, a 10-machine flow shop production line, a 50-kWh energy storage system, a 100-kW gas turbine,  and the grid for power supply, demonstrating the validity and applicability of this event-triggered hybrid scheduling (ETHS) framework.

\end{abstract}


\end{frontmatter}

\section{Introduction}
Due to the global energy crisis and 2050 net zero emission target set in Paris Agreement, renewable energy sources (RESs) such as solar energy, and  energy storage systems (ESSs) such as batteries are gradually adopted by the manufacturing industry. However, as pointed out in \cite{Impram_2020_Review}, RESs are hardly predictable, which makes the energy-aware scheduling in manufacturing systems extremely challenging to guarantee the production efficiency while improving the efficiency of RESs. 

Many techniques have been developed to improve the efficiency and optimality of energy-aware scheduling. These techniques can be classified into two major categories:  off-line optimisation (scheduling) and on-line scheduling.

Off-line scheduling provides an optimal planning for the production lines and energy management for a given cost function based the predicted performance using the various models of the production lines and RESs without using real-time observed data. In off-line scheduling, the choice of the cost function is one of the major considerations. For example, 
\cite{b2} highlighted the need of reducing the emission in the cost function along with the guarantee of the production throughput, leading to less utilisation rate of RESs. \cite{b3} utilised a cost function that can balance or coordinate production schedule and energy management so that the battery management and production schedule are optimised simultaneously. Similarly, \cite{b6} simulated a 3-machine job shop and optimise with a multi-objective function that the  balance between makespan \footnote{The makespan is the time difference between the start and finish of a sequence of jobs or tasks.} and cost related to the energy management.
Different types of RESs have been also considered. For example, a diesel generator was considered in \cite{b4}.  
As the off-line scheduling only runs once for the energy-aware scheduling, the computational cost is not an issue. Thus it can handle a very complex system with many constraints and tuning parameters for a long-term prediction. In the off-line scheduling, the prediction is based on the model of the production line, and RESs, the modeling uncertainties coming from the unpredictable nature of RESs or the unexpected machine failures of the production line are inevitable. In general, off-line scheduling is not robust with respect to these modeling uncertainties.

Compared with the model-based off-line scheduling, on-line scheduling also utilises the real-time data measured from various sensors equipped in green manufacturing systems. It consists of two different types of strategies. One is on-line adjustment to minimize the mismatch between the off-line scheduling and the measured current status in terms of the green manufacturing systems. For example, \cite{b7} compensated the mismatch between real-time RES supply and production demand from the off-line scheduling while \cite{b8} dealt with the mismatch in order to improve utilisation rate for RES. The other one is on-line rescheduling, in which the current measurements are used to predict the future performance based on the models of the production lines and RESs. On-line rescheduling depends on the the cost function and the re-scheduling rate.  For example, in \cite{b9}, a rescheduling happened hourly. The production cost and tardiness were considered in the selection of the cost function in \cite{b10}. 
As on-line measurements, which reflect the current status of the green manufacturing system, are used in the prediction along with the models, the on-line schedule is more robustness to modeling uncertainties compared with the off-line scheduling. However, both on-line adjustment and on-line rescheduling require an extra computational power to solve some scheduling problem. In particular, for a complex system with many tuning parameters and constraints and a longer prediction horizon, the computational cost becomes expensive. 


There are some attempts trying to combine off-line scheduling and on-line scheduling to balance the robustness with the modeling uncertainties and the computational cost. For example,  \cite{b12} suggested using the off-line scheduling as a reference for production while managing batteries in real-time to deal with the variance to the prediction. In \cite{b13}, production schedule was decided in off-line scheduling, while on-line adjustment focuses on dealing with difference between wind power prediction and observation. Similarly in \cite{b14}, grid procurement was considered as a tuning parameter in the on-line scheduling when using the reference coming from the off-line scheduling reference. These attempts are highly heuristic and case-dependent. A systematic way of combining off-line scheduling and on-line scheduling is still lacking.

This paper focuses on providing a systematic design framework for energy-aware scheduling in manufacturing systems based on the available imprecise models (off-line scheduling) and measurement data (on-line schedule). A so-called event-triggered hybrid scheduling (ETHS) is proposed based on a novel concept of partially dispatchable state, which can be served as an extra design freedom to balance the computational cost and the performance when the unexpected situations happen. When the observed performance is satisfying, no rescheduling is triggered to reduce the computational cost. When the observed performance is not satisfying, this partially dispatchable state can be triggered as a part of rescheduling to improve the performance. Such a systematic design is based on a rigorous mathematical problem formulation for off-line scheduling and on-line scheduling, to which are applicable to a large class of green manufacturing systems. 

This paper is organized as follows. Section 2 formulates both off-line scheduling and on-line scheduling, followed by the detailed design steps in the proposed ETHS in Section 3. A simulation example is presented in Section 4. Section 5 concludes this work. 

\section{Off-line scheduling and On-line scheduling in a manufacturing system}

This paper focuses on a class of manufacturing systems, of which consist a set of interconnected heterogeneous subsystems: a sequences of production lines, represented as $P$, some renewable energy systems (RESs), and some energy storage systems (ESSs). Each subsystem is characterized by a family of states and parameters. Usually, the state is characterized by a time-series sequence, which can be represented as $x[k]\in \CR$, $k\in\CN$ at the $k^{th}$ sampling instant, where $\CR$ is the set containing all real numbers and $\CN$ is the set containing all positive integers. For example, the state of charge (SOC) of the battery at $[k+1]^{th}$ time instant is related to SOC at $k^{th}$ time instant, power of charge and power of discharge at the $k^{th}$ time instant. Another example is the working condition of the production line $P$ at each sampling time instant in terms of $(on, off)$ of each machine is also a time series, it consists of on/off status of each machine at $k \in [0,N-1]$. 

It is expected that for each state $x[k]\in \CR, k\in\CN$, we have either a dynamic model or static model to characterize its behaviour over time. We care the performance of this manufacturing system within a finite time, i.e., $k\in [0,N-1]$. More precisely, it has
\bea
x[k+1]=f(x[k+1], x[k],\btheta_1), \hspace{0.2in} x[0]\in \CR, \label{state_equation}
\eea
for some nonlinear mapping $f:\CR\times \CR \times \CR^{n_\theta}$ is a known mapping with some parameters $\btheta_1 \in \CR^{n_\theta}$. When $f(x[k+1],x[k],\btheta)=x[k+1]+f_1(x[k],\btheta_1)$ for some nonlinear mapping $f(\cdot,\cdot)$, it becomes a static mapping. When $f(x[k+1],x[k],\btheta)=f_1(x[k],\btheta_1)$ holds, it becomes a dynamic system. It is noted that the nonlinear mapping $f(\cdot,\cdot,\cdot)$ can also represent the Boolean logic such as the machine is on ($1$) or off ($0$).  

Next, we categorize the state into two categories.


\begin{enumerate}

\item [1] Non-dispatchable state $\bfx_{ND}^n[k]\in \CR^{n_{ND}}$ for any $k\in [0, N-1]$ represents all the time-series in the manufacturing system, whose information can be observed but cannot be manipulated. For example, in a RES subsystem, which has a solar panel, the temperature and radiation at each sampling instant $k$ belong to non-dispatchable states. 

\item [2] Dispatchable state $\bfx_{D}[k]\in \CR^{n_{D}}$ for any $k\in [0,N-1]$ represents all the time-series in the manufacturing system, whose information can be both observed and manipulated. For example,  the charging and discharging sequence of a battery of an ESS subsystem are dispatchable states.  
\end{enumerate}
Similarly, we also classify the parameters involved in the manufacturing system into two categories.
\begin{enumerate}
    \item [1] Non-tunable parameters. The notation of $\btheta_{NT}\in \CR^{n_{\theta, NT}}$ represents a family of parameters in the manufacturing system that cannot be tuned. For example, the number of machine in the production line $P$ is a non-tunable parameter.
    \item [2] Tunable parameters. The notation of $\btheta_{T}\in \CR^{n_{\theta, T}}$ represents a family of parameters in the manufacturing system that can be participant into the scheduling as a part of scheduling parameters. 
\end{enumerate}

Letting $n_s=n_{ND}+n_{D}$ and $n_p=n_{\theta,NT}+n_{\theta,T}$, it follows that 
\bea
\overrightarrow\bfx[k]=\left[\begin{array}{cc}\bfx_{ND}^T[k]\ &\bfx_{FD}^T[k] \end{array}\right]^T\in \CR^{n_{s}},\hspace{0.2in}\label{all_state}
\eea
and
\bea
\overrightarrow\btheta=\left[\begin{array}{cc}\btheta_{NT}^T\ &\btheta_{T}^T\ \end{array}\right]^T\in \CR^{n_{p}},\hspace{0.2in}
\eea
the model to characterize $\overrightarrow\bfx[k]$ can be re-written as 
\bea
\Sigma_M: \overrightarrow\bfx[k+1]=\overrightarrow\bff\left(\overrightarrow\bfx[k+1], \overrightarrow\bfx[k],\overrightarrow\btheta\right), \  \overrightarrow \bfx[0]\in \CR^{n_s}\label{vector_state_equation}
\eea
where the nonlinear mapping $\overrightarrow\bff: \CR^{n_{s}}\times \CR^{n_{s}}\times \CR^{n_p}\rightarrow \CR^{n_{s}}$ is known. 

We also denote that the prediction using the model (\ref{vector_state_equation}) at the $s^{th}$ sampling instant from the measurements at the $k^{th}$ sampling instant as ${\overrightarrow {\hat \bfx}}[s|k]$ for any $s> k$ and $k\in [0,N-1]$. This leads to two different types of scheduling problems in literature: off-line scheduling/scheduling and on-line scheduling/scheduling.

With the notations introduced, next will provide a rigor mathematical formulation of off-line scheduling and on-line scheduling.

\subsection{Off-line scheduling}
There are many algorithms in literature, which can be formulated as an off-line scheduling. Denoting 
\beas 
   \overrightarrow {\hat \bxi}[N-1|0]=\left[\begin{array}{ccc} \overrightarrow {\hat\bfx}_{D}^T[1|0] &\cdots&\overrightarrow {\hat\bfx}_{D}^T[N-1|0]\end{array}\right]^T,
\eeas 
which is a vector in $\CR^{n_{off}}$ where $n_{off}=n_{D}\times (N-1)$, the off-line scheduling problems can be formulated as 
\bea
\min_{\overrightarrow {\hat \bxi}[N-1|0]\in \CR^{n_{off}}, \btheta_{T}\in \CR^{n_{\theta, T}}}\displaystyle \sum_{k=1}^{N}\bfl_{1,k}\left({\overrightarrow {\hat \bfx}}[k|0],\overrightarrow\btheta\right)\hspace{0.3in}\label{Cost_off_line}\\
s.t. \left\{\begin{array}{l}\overrightarrow\bfx[k+1]=\overrightarrow\bff\left(\overrightarrow\bfx[k+1], \overrightarrow\bfx[k],\overrightarrow\btheta\right), \overrightarrow\bfx[0]\in \CR^{n_s}\\
\bfg\left({\overrightarrow {\hat \bfx}}[k|0],\overrightarrow\btheta\right)=\bm{0}_{p_1} \\
\bfh\left({\overrightarrow {\hat \bfx}}[k|0],\overrightarrow\btheta\right)\geq \bm{0}_{p_2},\forall k=0,1,\ldots, N-1.
\end{array}
\right.,\label{off_line_scheduling}
\eea
where $\bfl_{1,k}(\cdot,\cdot)$ is the time-varying cost function with an appropriate dimension. The optimal solution of the off-line scheduling at the $k^{th}$ instant is denoted as $J_{off-line}^*[k]$, which is the computed optimal value of $\bfl_{1,k}$ when the dispatchable state takes the optimal value. 
%

It is highlighted that many existing algorithms in manufacturing systems with renewable resource management can be re-formulated as (\ref{off_line_scheduling}). For example, in \cite{b17}, machines of production lines, batteries, combined heat and power (CHP) plants were used to improve energy efficiency while keeping the production lines running properly by minimizing the makespan of a manufacturing system. The solutions of off-line scheduling can be found by using existing scheduling techniques such as discrete whale optimisation by \cite{b15} and particle swarm optimisation by \cite{b16}.

\begin{remark}
Due to the complexity of the manufacturing system in the presence of RESs and ESSs, the off-line scheduling with the cost (\ref{Cost_off_line}) and constraints (\ref{off_line_scheduling}) usually is quite complicated with hundreds of variables to tune and dozens of constraints. This is one of the major reasons that off-line scheduling is preferred as it can ignore the computational cost. The performance of the off-line scheduling is highly dependent on the accuracy of the model used, i.e., $\overrightarrow\bff(\cdot,\cdot,\cdot)$. If the model is not precise, the solutions obtained from the off-line scheduling is not optimal. Moreover, the modeling errors can be propagated over time, leading to undesirable performance. \hfill $\circ$
\end{remark}

In order to address the robustness of the off-line scheduling with respect to unmodelled uncertainties, \cite{b17} utilised the probability distribution of the machine breakdown, resulting in the stochastic cost function. The scheduling was thus performed by exploiting the statistic properties of this stochastic cost function such as its mean and variance. Although some of robust algorithms with respect to modeling uncertainties and unexpected running situation have been proposed, in general, the off-line scheduling lacks of robustness.

\subsection{On-line scheduling} 
On-line scheduling means that at each time instant, new measurements $\overrightarrow\bfx[k]\in \CR^{n_s}$ is added and used in scheduling. For the simplicity of the presentation, we define $N_s$ as the prediction horizon and the control horizon. In a more general setting, the prediction horizon and the control horizon can be different. For example, a 48-hour prediction horizon for electricity generation and 1-hour control horizon for energy management system was used in \cite{b20}. 
%
%
%
%
%
%
Denoting
\beas 
  &\ &\overrightarrow {\hat \bxi}_s=\overrightarrow {\hat \bxi}_s[k+N_s-1|k]
   \nonumber\\
   &=&\left[\begin{array}{ccc} \overrightarrow {\hat\bfx}_{D}^T[k+1|k] &\cdots&\overrightarrow {\hat\bfx}_{D}^T[k+N_s|k]\end{array}\right]^T,
\eeas 
which is a vector in $\CR^{n_{k,N_s}}$ where $n_{k,N_s}=n_{D}\times (Ns-1)$. The on-line scheduling can be formulated as
\bea
\min_{\overrightarrow {\hat \bxi}_s\in \CR^{n_{k,N_s}}, \btheta_{T}\in \CR^{n_{\theta, T}}}\displaystyle \sum_{s=0}^{N_s}\bfl_{2,s}\left({\overrightarrow {\hat \bfx}}[k+s|k],\overrightarrow\btheta\right)\hspace{0.2in}\label{Cost_on_line}\\
s.t. \left\{\begin{array}{l}\overrightarrow\bfx[k+1]=\overrightarrow\bff\left(\overrightarrow\bfx[k+1], \overrightarrow\bfx[k],\overrightarrow\btheta\right), \overrightarrow\bfx[k]\in \CR^{n_s}\\
\bfg\left({\overrightarrow {\hat \bfx}}[k+s|k],\overrightarrow\btheta\right)=\bm{0}_{p_1} \\
\bfh\left({\overrightarrow {\hat \bfx}}[k+s|k],\overrightarrow\btheta\right)\geq \bm{0}_{p_2},\forall s=0,\ldots, N_s.\label{on_line_scheduling}
\end{array}
\right.
\eea
where $\bfl_{2,s}(\cdot,\cdot)$ in (\ref{Cost_on_line}) can be also time-varying. It might be different from $\bfl_1(\cdot,\cdot)$ in (\ref{Cost_off_line}).

The mathematical description of on-line scheduling consisting of (\ref{Cost_on_line}) and the constraints (\ref{on_line_scheduling}) is very general. It contains both on-line adjustment without using prediction models and on-line rescheduling. When $N_s=0$, the on-line scheduling problem becomes an on-line adjustment. More precisely,  by using the current measurements $\overrightarrow\bfx[k]\in \CR^{n_s}$, it is possible to re-tune the dispatchable state $\overrightarrow {\hat \bxi}_0$ to reach the optimal solution of the cost function $\bfl_{2,0}(\cdot,\cdot)$. 

When $N_s\geq 1$, this problem becomes on-line rescheduling based on the predicted future behaviours using the model and the current measurements $\overrightarrow\bfx[k]$. In particular, when $N_s\geq 2$, the optimal solution has the predicted future $N_s$ solution, i.e.,  $\overrightarrow {\hat \bxi}_s^*=\overrightarrow {\hat \bxi}_s^*[k+N_s-1|k],N_s\geq 2$.  Under such a situation, the concept of receding horizon design and model predictive control (MPC) \cite{b19,b21}, in which only the first step of the predicted optimal solution, is used. 

There are many existing algorithms proposed for optimizing the production line performance in the presence of RESs. For example, in \cite{b18}, rescheduling was allowed in the flexible job scheduling by using the latest measurements, providing robustness and flexibility. In \cite{b19}, MPC was used to optimize the performance of renewable energy components such as batteries in microgrids.
\begin{remark}
Different from off-line scheduling, on-line scheduling includes on-line rescheduling, which utilizes the latest measurement $\overrightarrow\bfx[k]$ and the model to predict the future behaviour within a finite prediction horizon $N_s$. On one hand, such a setting leads to more robust optimal performance with respect to the modelling uncertainties by using the latest measurements. On the other hand, this receding horizon requires a rescheduling at each time step and greatly increases computational cost. \hfill $\circ$
\end{remark}

Both off-line scheduling and on-line scheduling techniques have their advantages and disadvantages in terms of computational cost and robustness with respect to modelling uncertainties. This work aims to balance the robustness, optimality, computational cost, and flexibility by providing a novel design framework, which introduces a new design freedom using the partially-dispatchable state, and integrating the off-line scheduling with an event-trigger on-line scheduling. This novel framework is called an event-triggered hybrid scheduling (ETHS).

\section{Event-triggered Hybrid scheduling}

For simplicity, in this section, we remove the design of the optimal tunable parameter $\btheta_T$ in the cost functions (\ref{Cost_off_line}) and (\ref{Cost_on_line}).  In order to increase the flexibility of the proposed ETHS, we further decompose of the dispatchable state into two sub-classes. 
One is called fully-dispatchable state $\bfx_{FD}\in \CR^{n_{FD}}$ and the other is called partially-dispatchable state  $\bfx_{PD}\in \CR^{n_{PD}}$. 
We use the fully-dispatchable state in both off-line and on-line scheduling.
The partially-dispatchable state will be used in off-line scheduling at the beginning. At the $k^{th}$ time instant, if the predicted performance is good enough, $\bfx_{PD}[k]$ will be treated as the non-dispatchable state, and will not be involved in on-line scheduling. If the predicted performance is not satisfying, $\bfx_{PD}[k]$ will be triggered as a fully-dispatchable state, participating into on-line rescheduling. 


The concept of partially-dispatchable state is introduced. More precisely, the partially-dispatchable state $\bfx_{PD}[k]$ can be represented as:
\bea 
\bfx_{PD}[k]=\left[\begin{array}{ccc} x_{PD,1}[k]&\ldots&x_{PD,n_{PD}}[k]\end{array}\right]^T\in \CR^{n_{PD}},\label{def_partially_dispatchable}
\eea
for any $k\in [0,N-1]$. The idea of using this novel partially dispatchable state is to balance the robustness with respect to modelling uncertainties and the computational cost. Moreover, choosing this partially dispatchable state becomes an extra design freedom in ETHS, providing more flexibility. This partially dispatchable state can be a non-dispatchable state in some sub-intervals of $[0,N-1]$ while in other sub-intervals of $[0,N-1]$, it becomes a fully-dispatchable state. For the convenience of the notation, it is denoted that $\Omega_{s,i}$ is union of time intervals within the interval $[s,N-1]$ in which the $i^{th}$ partially-dispatchable state $x_{PD,1}$ can be fully manipulated.

Consequently, the dispatchable state $\bfx_D$ can be re-written as
\bea
\bfx_{D}[k]=\left[\begin{array}{c}\bfx_{FD}^T[k]\\\bfx_{PD}^T[k]\end{array}\right]^T \in \CR^{n_{D}}.\label{dispatchable_decompose}
\eea

The proposed ETHS consists of four steps (S1--S4) as shown in Figure 1.

\begin{figure}[htb!]

\includegraphics[width=9cm]{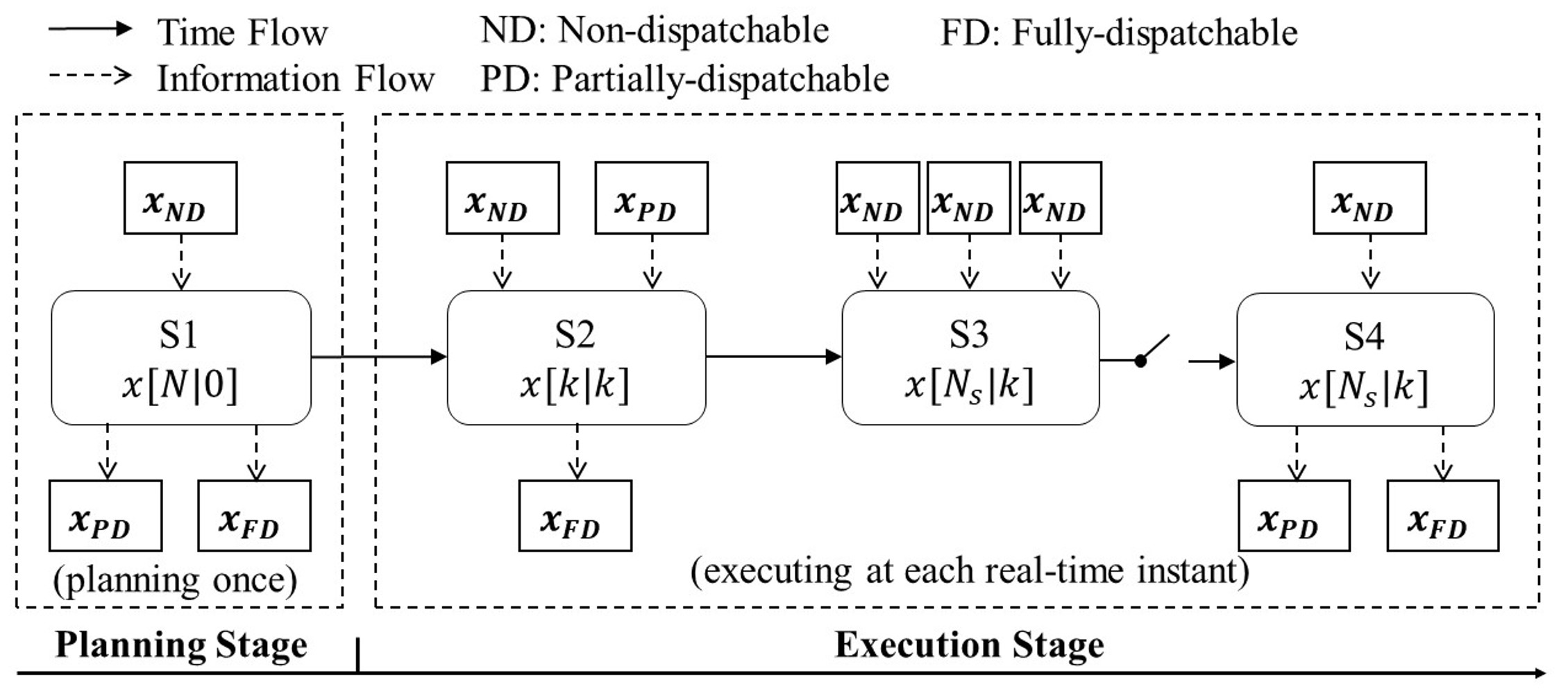}    

\caption{The diagram of the event-triggered hybrid scheduling} 

\label{fig:diagram}

\end{figure}
\begin{enumerate}
\item [S1]: Off-line scheduling defined by the cost (\ref{Cost_off_line}) and constraints (\ref{off_line_scheduling}) to find an off-line optimal state 
\bea
\overrightarrow {\hat \bxi}^*[N-1|0]=\left[\begin{array}{c}\left(\overrightarrow {\hat \bxi}_{FD}^*[N-1|0]\right)^T\\ \left(\overrightarrow {\hat \bxi}_{PD}^*[N-1|0]\right)^T\end{array}\right]^T, \label{optimal_solution_off_line}
\eea
\item [S2]: On-line adjustment for fully-dispatchable state $\bfx_{FD}^T[k]$. More precisely, at the each step, solve the on-line scheduling problem (\ref{Cost_on_line}) and the constraints (\ref{on_line_scheduling}) with the scheduling parameter of fully-dispatchable state $\bfx_{FD}[k]$ with $N_s=0$ while the partially-dispatchable state is a constant vector coming from $\overrightarrow {\hat \bxi}_{PD}^*[N-1|0]$ in (\ref{optimal_solution_off_line}). The solution of the on-line adjustment is $\bfx_{FD}^*[k]$.

\item [S3]: Evaluating the predicted future performance using the current measurements of $\bfx_{ND}[k]$ and optimal fully-dispatchable state $\bfx_{FD}^*[k]$ while the partially-dispatchable state $\overrightarrow {\hat \bxi}_{PD}^*[N-1|0]$ is treated as a constant. For the convenience of notation, we denote
\bea
\overrightarrow\bfz[k]=\left[\begin{array}{c} \bfx_{FD}^T[k]\\
\bfx_{ND}^T[k]\end{array}\right]^T,
\eea
and the parameter $\balpha=\overrightarrow {\hat \bxi}_{PD}^*[N-1|0]$ coming from S2. Moreover, we denote 
\bea
\overrightarrow {\bar \bfz}[k]=\left[\begin{array}{c} \left(\bfx_{FD}^*[k]\right)^T\\
\bfx_{ND}^T[k]\end{array}\right]^T,\label{bar_z_initial}
\eea
consequently, the future performance in terms of the following cost function is evaluated:
\bea
{\bf J}[k]=\displaystyle \sum_{s=1}^{N-k+1}\bfl_{3,s}\left(\overrightarrow {\bar \bfz}[k+s|k],\overrightarrow\btheta,\balpha\right)\hspace{1,0in}\label{performance}\\
s.t. \overrightarrow {\bar \bfz}[k+1]=\overrightarrow\bff_z\left(\overrightarrow{\bar \bfz}[k+1], \overrightarrow {\bar \bfz}[k],\btheta\right), \hspace{0.2in}\nonumber
\eea
with the initial condition at  $\overrightarrow {\bar \bfz}[k]$. Here $\overrightarrow {\bar \bfz}[k]$ is defined in (\ref{bar_z_initial}) and $\overrightarrow\bff_z$ is some parts of $\overrightarrow\bff(\cdot,\cdot,\cdot)$ defined in (\ref{off_line_scheduling}) or (\ref{on_line_scheduling}). Here the nonlinear mapping $\bfl_{3,k}(\cdot,\cdot,\cdot)$ can be different from $\bfl_1$ in (\ref{Cost_off_line}) or $\bfl_2$ (\ref{Cost_on_line}). It is assumed that ${\bf J}[k]\in \CR^{n_J}$.

\item [S4]: Trigger on-line re-scheduling if needed. Precisely, if the performance ${\bf J}[k]$ is not satisfying, i.e., if
\bea
\ell\left({\bf J}[k]\right)\geq \varepsilon, \label{threshold}
\eea
where $\ell(\cdot):\CR^{n_J}\rightarrow \CR_{\geq 0}$ and $\varepsilon$ is a pre-defined positive constant, then on-line rescheduling will be triggered, hence it will solve the on-line scheduling problem (\ref{Cost_on_line}) and the constraints (\ref{on_line_scheduling}) with the scheduling parameter of the dispatchable state $\bfx_{D}[k]$ with $N_s\geq 1$.
\end{enumerate}

It is noted that both off-line scheduling (in S1) and on-line scheduling (in S2 and S4 when the rescheduling is triggered) are performed. The performance of on-line adjustment is evaluated using the model to decide whether a rescheduling is needed. Different from the existing off-line scheduling and on-line scheduling, the number of manipulated variables changes at different stage due to the introduction of the partial dispatchable state.

\section{A Simulation Example}

This section shows how to use the proposed event-triggered hybrid scheduling framework to design a flow shop manufacturing system integrated with PV so as to reduce energy costs and guarantee production throughput. This section starts from a brief description of the system, followed by a few scenarios to address the following questions in designing an ETHS: 
\begin{enumerate}
    \item How to select partially-dispatchable state?
    \item How to implement on-line adjustment?
    \item How to select the evaluating function ${\bf J}[k]$ in (\ref{performance})?
    \item How to select event-trigger function $\ell(\cdot)$ and the threshold $\varepsilon$ in (\ref{threshold})?
\end{enumerate}
These scenarios will provide design guidelines for engineering practitioners and demonstrate the validity, applicability and flexibility of the proposed framework.

\subsection{Description of the System}
This subsection provides a description of the system, including the system components, data sources, operational task, randomness during simulation, and all states and parameters.

The manufacturing system consists of a 100-kW PV system, a 10-machine flow shop production line, a 50-kWh Energy Storage System (ESS), a 100-kW gas turbine (GaT), and the grid for power supply and surplus power feed-in. Solar irradiance data are from the Australian Bureau of Meteorology, grid electricity price and feed-in tariff are from SUMO, and flow shop machine power consumption and operation time are acquired by measuring and scaling up the empirical data from factories. The example is to operate the introduced system by deciding each component's states to finish 25 jobs and reduce energy cost for 24 hours with a time resolution of 5 minutes, namely 288 time steps. The energy flow and material flow can be found in Figure \ref{fig:diagram}.

\begin{figure}[htb!]

\includegraphics[width=9cm]{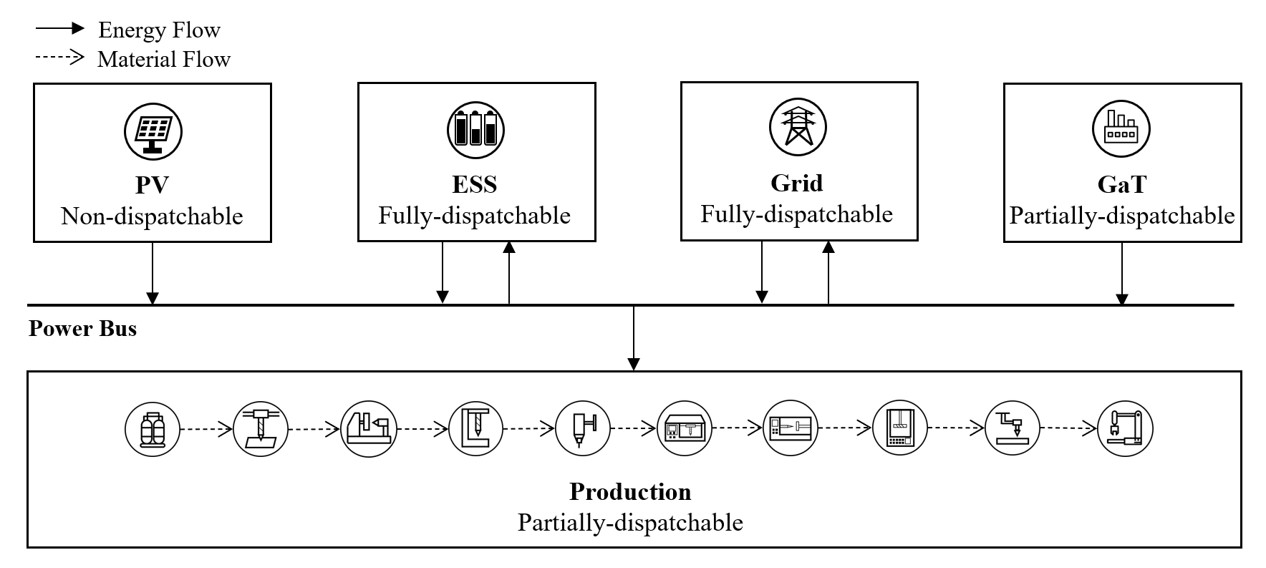}    

\caption{Energy Flow and Material Flow of the Manufacturing System} 

\label{fig:diagram}

\end{figure}

In the simulation, the machine breakdown processes some randomness. The possibility of breakdown and the time when the breakdown happens are assumed to follow an exponential distribution, which was proposed by \cite{He_Sun_2013,Tang_Dai_Salido_Giret_2016}. For 10 machine states defined as $\bfx_2$ (see Table \ref{table_state} for the definition of states), it has the form of
\bea
\arraycolsep=1.4pt\def\arraystretch{1.5}
\left\{\begin{array}{l}
P(\hat{\bfx}_{2,i}[k]=0|\bfx_{2,i}[k]=1)=1-e^{-\theta_{12}T_{on}}\\
P(\hat{\bfx}_{2,i}[k+1]=1|\hat{\bfx}_{2,i}[k]=0)=1-e^{-\theta_{13}T_{bd}}
\end{array}\right., \label{breakdown}
\eea
where $T_{on}$ is the accumulated operating time for a machine since the last observed breakdown $\hat{\bfx}_{2,i}[k]$, $T_{bd}$ means the accumulated observed breakdown at the next time step based on the current breakdown. Here $\theta_{12}$ and $\theta_{13}$ are parameters used to characterize the probability distribution (see details in Table 2).


In this simulation, the initial selection of states is listed in Table \ref{table_state}. For example, $x_1$ is the state variable for PV, the value of $x_1[k]$ means the power of PV generation at the $k^{th}$ sampling instant. When a vector state such as $\bfx_4\in \CR^{10}$ is used, $x_{4,i}[k]$ means its $i^{th}$ element at the $k^{th}$ sampling instant. The similar method is used for defining parameters, which are summarized in Table 2. As discussed in Section 3, all the parameters are non-tunable. 

\begin{table}[htb!]
\caption{States Variables for the Simulation}
\label{table_state}
\resizebox{\columnwidth}{!}{%
\begin{tabular}{|l|l|}
\hline
States & Description \\ \hline
ND States &  \\ \hline
$x_1 \in \CR^{1}$ & Power of PV generation \\ \hline
PD States &  \\ \hline
$\bfx_2 \in \CR^{10}$ & Machine status on or off for each machine \\ \hline
$\bfx_3 \in \CR^{10}$ & Operation status starts or not start on each machine \\ \hline
$\bfx_4 \in \CR^{10}$ & Number of operations finished on each machine \\ \hline
$x_5 \in \CR^{1}$ & GaT generation power \\ \hline
FD States &  \\ \hline
$x_6 \in \CR^{1}$ & ESS charging power \\ \hline
$x_7 \in \CR^{1}$ & ESS discharging power \\ \hline
$x_8 \in \CR^{1}$ & ESS state of charge (SOC) \\ \hline
$x_9 \in \CR^{1}$ & Grid electricity procurement power \\ \hline
$x_{10} \in \CR^{1}$ & Grid electricity feed-in power \\ \hline
\end{tabular}
}
\end{table}

\begin{table}[htb!]
\caption{Parameters for the Simulation}\label{table_parameter}
\resizebox{\columnwidth}{!}{%
\begin{tabular}{|l|l|l|} 
\hline
Parameters & Description & Value \\ \hline
$\btheta_1 \in \CR^{10}$ & \begin{tabular}[c]{@{}l@{}}Machine power for \\ 10 machines (kW)\end{tabular} & \begin{tabular}[c]{@{}l@{}}[50.63,22.4,5.12,5.28,\\ 12.68,35.14,6.58,\\ 5.21,8.4,2.15]\end{tabular} \\ \hline
$\btheta_2 \in \CR^{10}$  & \begin{tabular}[c]{@{}l@{}}Operation time for \\ 10 operations (5 min)\end{tabular} & [5,2,8,5,3,8,4,6,6,7] \\ \hline
$\theta_3 \in \CR^{1}$  & Gas price (\$) & 1.83 \\ \hline
$\theta_4 \in \CR^{1}$  & \begin{tabular}[c]{@{}l@{}}ESS charging/discharging\\ efficiency\end{tabular} & 0.9 \\ \hline
$\theta_5 \in \CR^{1}$  & \begin{tabular}[c]{@{}l@{}}ESS fixed operational \\ cost (\$/5 min)\end{tabular} & 0.003 \\ \hline
$\theta_6 \in \CR^{1}$  & \begin{tabular}[c]{@{}l@{}}ESS charging/discharging \\ degradation cost (\$/kWh)\end{tabular} & 0.0006 \\ \hline
$\theta_7 \in \CR^{1}$  & ESS maximum capacity (kWh) & 50 \\ \hline
$\theta_8 \in \CR^{1}$  & ESS depth of discharge & 80\% \\ \hline
$\theta_9 \in \CR^{1}$ & \begin{tabular}[c]{@{}l@{}}ESS maximum charging/\\ discharging power (kW)\end{tabular} & 50 \\\hline
$\btheta_{10} \in \CR^{288}$  & \begin{tabular}[c]{@{}l@{}}Grid electricity \\ procurement price (\$/kWh)\end{tabular} & \begin{tabular}[c]{@{}l@{}}0.330 from 9am to 9pm\\ 0.187 from 9pm to 9am\end{tabular} \\ \hline
$\theta_{11} \in \CR^{1}$  & \begin{tabular}[c]{@{}l@{}}Grid electricity \\ feed-in tariff (\$/kWh)\end{tabular} & 0.052 \\ \hline
$\theta_{12} \in \CR^{1}$  & \begin{tabular}[c]{@{}l@{}}Machine breakdown  \\ possibility parameter\end{tabular} & 0.002 \\ \hline
$\theta_{13} \in \CR^{1}$  & \begin{tabular}[c]{@{}l@{}}Machine breakdown time\\ parameter\end{tabular} & 1 \\ \hline
$\theta_{14} \in \CR^{1}$  & Number of production task & 25 \\ \hline 
\end{tabular}%
}

\end{table}

\subsection{Off-line Scheduling, On-line Scheduling and ETHS}

This subsection will compare the performance among off-line scheduling, on-line schedule and the proposed ETHS. Moreover, the tuning freedom in the proposed ETHS will be discussed. 


\noindent {\underline {Off-line scheduling}} 

Off-line scheduling is conducted with the objective function of minimising total operational cost $\bfl_{1,k}$ in (\ref{off_line_scheduling}), defined as follows:
\bea
\begin{split}
\bfl_{1,k}=&\bfx_9[k]\cdot \btheta_{10,k}-\bfx_{10}\cdot \theta_{11}+\theta_5+\\ 
&(\bfx_6[k]+\bfx_7[k])\cdot \theta_4+\bfx_5[k]\cdot\theta_3
\end{split}\label{simulation_off_line_objective}
\eea

The constraints used in this simulation include the state constraints at each sampling instant namely for any $k=1,2,\ldots, 288$:
\bea
\left\{\begin{array}{l}
\bfx_6[k] \leq \theta_9 \\
\bfx_7[k] \leq \theta_9 \\
\bfx_8[k] \geq \theta_7 \cdot (1-\theta_8)/2\\
\bfx_8[k] \leq \theta_7 \cdot (1+\theta_8)/2\\
\bfx_8[k+1]=\bfx_8[k]+\theta_4\cdot\bfx_6[k]-1/\theta_4\cdot\bfx_7[k]\\
\phi_1[k]=
\displaystyle\sum_{i=1}^{10} \bfx_{2,i}[k]\cdot\btheta_{1,i} + \bfx_{10}[k] + \bfx_7[k]
\end{array}
\right.\label{simulation_off_line_state}
\eea
where $\phi_1[k]=\bfx_1[k]+\bfx_9[k]+\bfx_7[k] +\bfx_5[k]$. There are also terminal constraints at the starting point or the end point of $288$ sampling points:
\bea
\left\{\begin{array}{l}
\bfx_8[0]=\theta_7\cdot0.5\\
\bfx_8[288]=\theta_7\cdot0.5\\
\bfx_{4,10}[288] \geq \theta_{14} \\
\end{array}
\right.\label{simulation_off_line_terminal}
\eea
Finally are the constraints coming from the flow shop. For the $i^{th}$ job, $i=1,\ldots, 10$, it has the following constraints:
\bea
\left\{\begin{array}{l}
\\
\bfx_{4,i}[k]=0, k=1,2,\ldots, \btheta_{2,i}.\\
\bfx_{4,i}[k]=\displaystyle\sum_{j=0}^{k-\theta_{2,i}} \bfx_{3,i}[j], k=\btheta_{2,i},\ldots,288 \\
\bfx_{2,i}[k] =\displaystyle \sum_{j=0}^{k} \bfx_{3,i}[j], k=1,2,\ldots,\btheta_{2,i}\\
\bfx_{2,i}[k-1] = \displaystyle \sum_{j=k-\theta_{2,i}}^{k} \bfx_{3,i}[j], k=\btheta_{2,i},\ldots,288\\
\bfx_{3,i}[k]\cdot\btheta_{2,i} \leq \displaystyle\sum_{j=k}^{k+\theta_{2,i}} \bfx_{2,i}[j], k=1,2,\ldots,288-\btheta_{2,i}\\
\end{array}
\right.\label{simulation_off_line_flow_shop}
\eea
Moreover, $\bfx_4$ has the dynamics, leading to the following constraint for $i=1,\ldots,9$
\bea
\bfx_{4,i}[k] \geq \bfx_{4,i+1}[k] + \bfx_{2,i+1}[k], k=1,2,\ldots,288. \label{simulation_off_line_battery}
\eea
These constraints cover the dynamic characterization for the state of ESS, flow shop assumptions, and energy balance equation, whose detailed explanations can be found in \cite{b9,b15}. 

\noindent {\underline {On-line scheduling}} 

On-line scheduling follows the objective and constraints in (\ref{Cost_on_line}-\ref{on_line_scheduling}). Specifically, we choose $\bfl_{2,s}=\bfl_{1,k}$, which means on-line scheduling shares the same objective with off-line scheduling. The length of time sequence is $N_s=288-k$, which means the cost compute from the current time until the end of day. Most of the constraints in on-line scheduling remain the same as off-line scheduling, except that we introduce a new design freedom $d$ to handle machine breakdown. 
This constraint is a terminal constraint in (\ref{simulation_off_line_terminal}), which indicates that the finished jobs at the end of the day should not be less than the production task. With the consideration that the machine breakdown might lead to unfinished jobs, in on-line scheduling this constraint is modified as $\bfx_{4,10}(288-k-d) \geq \theta_{14}$. Here $d$ is the buffer to finish the job. It can be time-varying. 

In this simulation, we design $d$ according to the accumulated machine breakdown time. The longer breakdown lasts, the more likely jobs are unfinished. Consequently, the remaining jobs should be accelerated. This leads to an earlier deadline $288-k-d$.


\noindent {\underline {Event-triggered hybrid scheduling}} 

ETHS follows the 4 steps presented in Section 3.

At S1, an off-line scheduling is conducted as (\ref{simulation_off_line_objective}-\ref{simulation_off_line_battery}). 

We define a new sequence to remember when the re-scheduling happens. When the re-scheduling happens at the $k^{th}$ sampling, it is denoted that $\tau_i=k$, $i=1,\ldots,$. 
We define $C[\tau_i]=J_{off-line}[\tau_i]^*$ as the scheduled cost.

At S2, an on-line adjustment is implemented as (\ref{Cost_on_line}) with $\bfl_{2,k}$ is defined as 
\bea
\begin{split}
\bfl_{2,k}=&\bfx_9[k]\cdot \btheta_{10,k}-\bfx_{10}\cdot \theta_{11}+\theta_5+\\ 
&(\bfx_6[k]+\bfx_7[k])\cdot \theta_4+\hat{\bfx}_5[k]\cdot\theta_3-C[k].\label{simulation_on_line_adjustment_objective}
\end{split}
\eea
The objective in (\ref{simulation_on_line_adjustment_objective}) is to  minimize the gap between the measured cost and the scheduled cost $C[k]$. The constraints are selected as
\bea
s.t. \left\{\begin{array}{l}
\bfx_6[k] \leq \theta_9\\
\bfx_7[k] \leq \theta_9\\
\bfx_8[k] \geq \theta_7 \\
\bfx_8[k] \leq \theta_7 \\
\phi_2[k]=\displaystyle\sum_{i=1}^{10} \hat{\bfx}_{2,i}[k]\cdot\btheta_{1,i} + \bfx_{10}[k] + \bfx_7[k].
\end{array}
\right.,\label{simulation_on_line_adjustment}
\eea
where $\phi_2[k]=\hat{\bfx}_1[k]+\bfx_9[k]+\bfx_7[k] +\hat{\bfx}_5[k]$. It can be seen that all non-dispatchable states and partially-dispatchable states become observed values, while fully-dispatchable states are control variables.

At S3, an evaluation function $J[k]$ is selected as 
\bea
\begin{split}
{\bf J}[k]=&\displaystyle \sum_{s=0}^{N_s}|\hat{x}_1[k+s|k]-x_1[k+s|k]| + \\\displaystyle &\sum_{i=1}^{10}\displaystyle \sum_{j=0}^{k}(1-\hat{\bfx}_{2,i}[j]), \label{jk}
\end{split}
\eea
 which accumulates the solar prediction error and the historical machine breakdown time. This cost indicates that the rescheduling will only be triggered when the solar prediction is inaccurate enough or the machine breakdown time is too long.

At S4, we choose $\ell\left({\bf J}[k]\right)={\bf J}[k]$, and we select the threshold as $\varepsilon=70$ in order to decide whether on-line rescheduling will be triggered or not.

Simulations results are summarized in Table 3 and Figure 3. %
With the possibility that some jobs are not able to finish, total energy cost becomes unfair to compare different scheduling techniques. Here the energy cost per job is selected as one of performance indices to compare three different scheduling schemes.  The simulation results show that off-line scheduling cannot handle unexpected events, resulting in unfinished jobs due to random machine breakdown while the computational cost in terms of computing time is low as it runs once every 24 hours. On-line scheduling has the best energy cost performance and is able to finish all jobs while it results in very large computational cost due to  rescheduling frequently.  It is highlighted that the proposed ETHS provides a good balance between the performance in terms of the energy cost per job and the computational cost. It is able to finish all jobs in the presence of random machine breakdowns with a reasonable computational cost. Moreover, the flexibility of fully-dispatchable components is highly utilised to deal with uncertainties coming from PV generation or random  machine breakdowns.
\begin{table}[h]
\caption{Scheduling Methods Comparison}\label{tb:margins}
\resizebox{\columnwidth}{!}{%

\begin{tabular}{cccc}

 & Finished Jobs & Energy Cost Per Job & Computingn Time/(s) \\ \hline
Off-line & 23.7 & 7.89 & 13 \\
On-line & 25 & 6.51 & 4604 \\ 
ETHS $\varepsilon=70$  & 25 & 7.10 & 508 \\ \hline
\end{tabular}%
}

\end{table}

\begin{figure}[htb!]

\includegraphics[width=9cm]{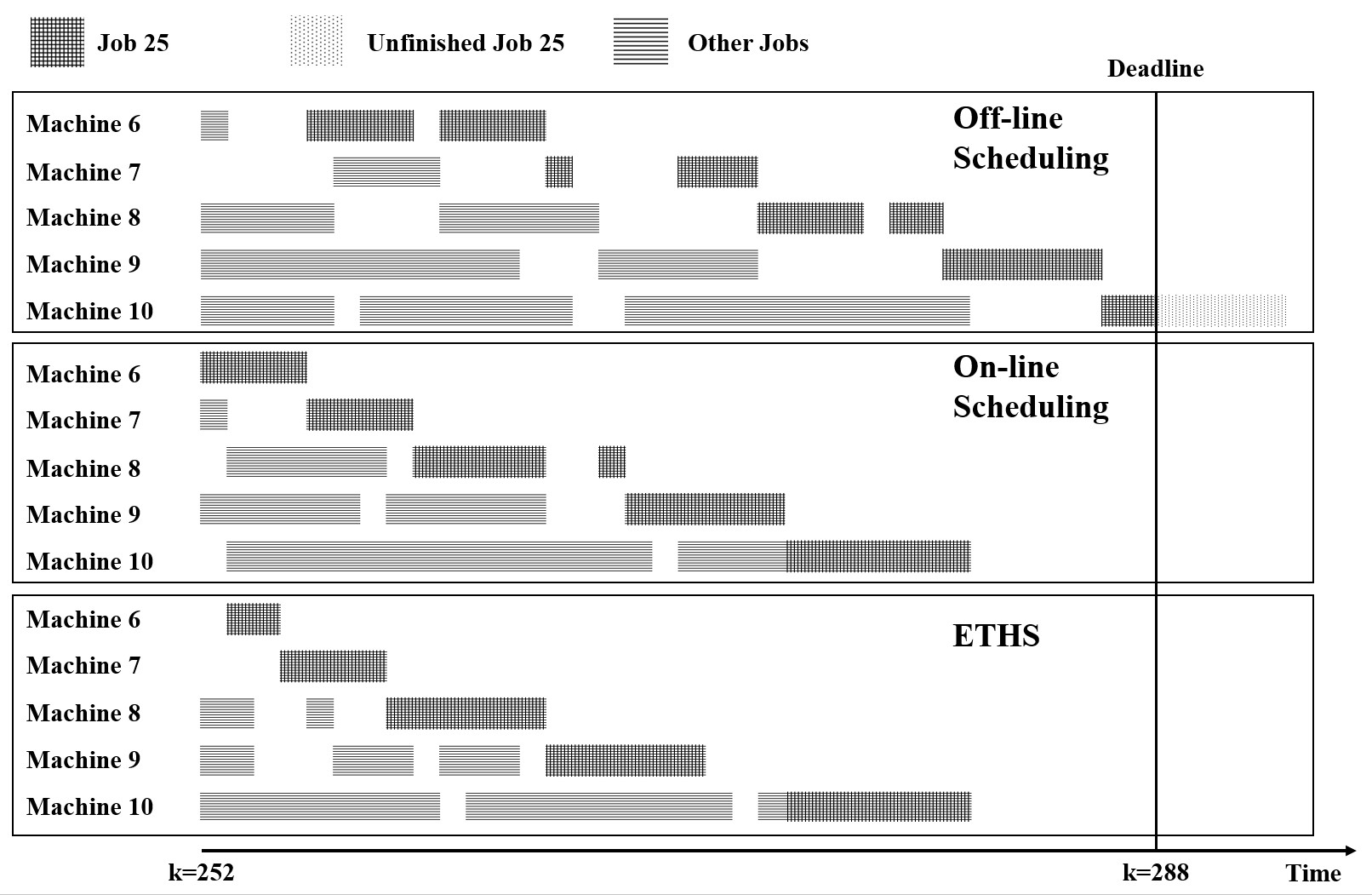}    

\caption{Flow Shop Gantt Charts Comparison} 

\label{gantt}

\end{figure}

\subsection{The Choices of Parameters of ETHS}

This subsection demonstrates the flexibility of the proposed ETHS by showing how different choices of parameters in four steps will affect its performance as four questions listed.  

\noindent {\underline {The choice of partial dispatchable state}}

Firstly, the categorisation of system components should consider both the physical property and the operational requirements. Fully-dispatchable components have higher flexibility than partially-dispatchable components, here we compare the performances of different categorisations in Table 4.

\noindent {\underline {Implementation of on-line adjustment}}

Secondly, to emphasise the importance of S2, we compare the performances under scenarios with or without S2. On-line adjustment targets at handling unserious uncertainty by utilising the high flexibility of fully-dispatchable components. Therefore, PV generation and demand energy errors between prediction and observation can be handled and the energy cost is reduced by 14.2\%, as shown in Table 4.

\noindent {\underline {The selection of the evaluation function}}

Thirdly, S3 to select $J[k]$ is decides whether it's necessary to trigger on-line rescheduling. Without the evaluation, the rescheduling will be executed at each time instant, resulting in better energy cost performance but much higher computing time, as shown in Table 4.
\begin{table}[h]
\caption{Parameter Selection Comparison}
\resizebox{\columnwidth}{!}{%
\begin{tabular}{lllll}
 & Scenario & \begin{tabular}[c]{@{}l@{}}Finished\\ Jobs\end{tabular} & \begin{tabular}[c]{@{}l@{}}Energy Cost\\ Per Job\end{tabular} & \begin{tabular}[c]{@{}l@{}}Computing\\ Time (s)\end{tabular} \\ \hline
\multirow{3}{*}{\begin{tabular}[c]{@{}l@{}}PD\\ Selection\end{tabular}} & Production, GaT & 25 & 7.10 & 508 \\
 & Production & 25 & 6.81 & 524 \\
 & None & 25 & 6.51 & 4604 \\ \hline
\multirow{2}{*}{\begin{tabular}[c]{@{}l@{}}On-line\\ Adjustment\end{tabular}} & With S2 & 25 & 6.51 & 4592 \\
 & Without S2 & 25 & 8.27 & 4604 \\ \hline
\multirow{2}{*}{\begin{tabular}[c]{@{}l@{}}Evaluation\\ Function\end{tabular}} & With S3 & 25 & 7.10 & 508 \\
 & Without S3 & 25 & 8.27 & 4604 \\ \hline
\end{tabular}%
}
\end{table}

\noindent {\underline {The choice of the threshold}}

Finally, the event-trigger threshold influences the frequency of rescheduling, thus resulting in different performances and computing time, as shown in Table 5.

\begin{table}[h]
\caption{Threshold Comparison}
\resizebox{\columnwidth}{!}{%
\begin{tabular}{ccccc}
Threhold  & Finished Jobs & Energy Cost Per Job & Computingn Time/(s) & \begin{tabular}[c]{@{}c@{}}Reschedule\\ Frequency\end{tabular} \\ \hline
$\varepsilon=30$ & 25 & 6.97 & 556 & 14 \\
$\varepsilon=70$ & 25 & 7.10 & 508 & 5 \\
$\varepsilon=100$ & 25 & 7.77 & 490 & 2 \\ \hline
\end{tabular}%
}
\end{table}


\section{Conclusion}
This paper proposed a novel event-triggered hybrid scheduling (ETHS) to balance the robustness, performance and computational cost in the energy-aware scheduling for the manufacturing systems. By mathematically formulating the existing off-line scheduling and on-line scheduling techniques, the proposed ETHS utilized the so-called partially-dispatchable state, which can be real-timely rescheduled when the desired performance is not satisfied. 
Compared with traditional off-line and on-line scheduling method, ETHS identifies the flexibility of heterogenous components in the manufacturing system, thus yielding outstanding performance with an acceptable computational cost. The flexibility of the 4-step ETHS was also demonstrated by a simulation example. The proposed ETHS can be extended to more complicated energy-aware manufacturing system beyond the flow shop.

\bibliography{main}             

\end{document}